\documentclass[twoside]{article}
\usepackage{graphicx}

\begin{document}
\title{Simulation of percolation on massively-parallel computers}
\author{Daniel Tiggemann\\Institute for Theoretical Physics, Universit\"at zu K\"oln\\50937 K\"oln, Germany, European Union\\Email: dt@thp.uni-koeln.de}
\date{\today}
\maketitle

\begin{abstract}
A novel approach to parallelize the well-known Hoshen-Kopelman
algorithm has been chosen, suitable for simulating huge lattices in high
dimensions on massively-parallel computers with distributed memory and message
passing. This method consists of domain decomposition of the
simulated lattice into strips perpendicular to the hyperplane of
investigation that is used in the Hoshen-Kopelman algorithm.
Systems of world record sizes,
up to $L=4000256$ in two dimensions, $L=20224$ in three, and $L=1036$ in four,
gave precise estimates for the Fisher exponent $\tau$, the corrections
to scaling $\Delta_1$, and for the critical number density $n_c$.
\end{abstract}

Percolation is a thoroughly studied model in statistical physics
\cite{stauffer}. The
algorithm invented by Hoshen and Kopelman in 1976 allows for examining large
percolation lattices using Monte Carlo methods \cite{hk}. Unfortunately, this
algorithm was invented for traditional sequential computers, and implementation
on modern parallel computers is far from trivial.

The normal way to parallelize an algorithm which works on regular data 
structures is domain decomposition. In this case, the investigated lattice is
cut into strips, and each processor is assigned one such strip for
investigation. As the Hoshen-Kopelman algorithm investigates the lattice
hyperplane by hyperplane (line by line in two dimensions; plane by plane in
three dimensions), various ways of domain decomposition can be characterized
by this hyperplane: for example, strips parallel or perpendicular to that
plane.

The easiest way would be to choose parallel strips, as in that case the
simulation of each strip can be carried out locally on each processor, only
after that communication is neccessary (exchanging the borders). Unfortunately,
this means that each processor has to store a full hyperplane of $L^{d-1}$
sites for a $L^d$ system. In two dimensions, this no problem at all, but in
higher dimensions this limits the size of systems that can be simulated.

When decomposing the system in perpendicular strips, each processor has to
store only a part of the hyperplane, and thus larger system sizes can be
simulated. Unfortunately, this advantage comes at a price: Frequent
interaction between neighbouring strips is neccessary throughout the
simulation, thus imposing a burden on runtime and complicating drastically
the implementation.

This latter approach was chosen for a masters thesis \cite{thesis}; a shortened
version will be presented here.

\section{Parallelizing Hoshen-Kopelman}
\noindent
The main problem when decomposing the lattice into perpendicular strips is
to cope with the non-regular communication patterns arising from the
Hoshen-Kopelman algorithm.

\subsection{Local and global clusters}
\noindent
Cluster that are confined within a strip and are not in touch with interfaces
between strips are called local; clusters that extend over several strips
are called global.

The Hoshen-Kopelman algorithm describes clusters by labels. In the sequential
version, a label can be a root label, describing a real cluster, or a 
non-root label pointing directly or indirectly to a root label (such labels
are generated when during the counting a new cluster seems to emerge, but later
is discovered to be part of another cluster).

For the parallel version, it is straightfoward to introduce three types of
labels: non-root labels, root labels associated with local clusters (local
root labels), and root labels associated with parts of global clusters 
(global root labels).

Associated with local root labels are the number of sites within the
corresponding cluster; global root labels carry the number of sites in the
cluster within the strip, and additionally pointers to the left and right
neighbour labels.

During the simulation, a local label can be changed to global, when it extends
to the interfaces of the strip. A global label can be changed to local
during the recycling process, when neighbouring labels are discarded.

When two previously different global clusters join at one site, the neighbouring
strips have to be informed of that fact, in order to join parts of the
corresponding cluster into one. This is called pairing, as pairs of labels
are joined. To reduce the number of messages that have to be sent, we gather
these and exchange information after the local part.

\subsection{Information exchange}
\noindent
After the local part within each strip has been simulated, communication
between the processors takes place to find out if clusters extend over the
interfaces, and what global labels are interconnected with each other.

\subsection{Recycling}
\noindent
An important step for economic memory usage is the frequent recycling of
non-root labels, as these carry no information, but are just an artefact of
the algorithm that speeds up computation. Additionally, also root labels
can be recycled, when they are not present in the current hyperplane of
investigation. These methods are known as Nakanishi recycling \cite{nakanishi}.

In the parallel implementation, we can use the same technique for purely
local labels; on the other hand,
we have to be careful not to recycle
labels that are still referenced by global labels in other strips.
We do that by exchanging information about these references; during this
process, references to non-root labels can be reclassified. In that way,
we can delete all non-root labels, even those pointing to global root labels.

Global root labels are recycled using a stepwise reduction of global clusters
along the strip: the rightmost strip that carries a part of a global cluster
checks if it is still alive; if not, it is recycled and the left neighbour
is informed that its right partner has vanished. If the left neighbour
has itself no left neighbour, it is converted to a local label.

\subsection{Step-by-step description of algorithm}
\noindent
The following list is a semi-formal description of the algorithm. Local
and communication part are repeated for each hyperplane the system consists
of, recycling is done whenever neccessary after the local and communication
part, and counting is done after the full system was examined.

\begin{enumerate}
\item \emph{Initialization}: Occupy the zeroth plane for busbar, if desired;
      initialize all data structures; etc.
\item \emph{Local}:
      \begin{enumerate}
      \item Examine the strip site by site. Do labeling.
      \item When two different global clusters join at one site, generate
            pairing information for left and right neighbour, but defer
            communication until after the local part.
      \end{enumerate}
\item \emph{Communication}:
      \begin{enumerate}
      \item Exchange borders of strip with neighbours.
      \item When two clusters of both strips join, convert clusters to global.
            If they are already global, but not yet connected, generate
            pairing information.
      \item Exchange pairing information. Pair global labels that belong
            together. During this, new pairing information can come up.
      \item Check if recycling is neccessary due to tight memory conditions.
      \end{enumerate}
\item \emph{Recycling} (if neccessary):
      \begin{enumerate}
      \item Reclassify the current hyperplane with root labels.
      \item Delete all non-root labels that point to local root labels.
      \item Reclassify the pointers to left and right of the global root
            labels by asking the neighbours for the corresponding root labels.
      \item Delete all remaining non-root labels.
      \item Mark all living local root labels and delete the non-marked ones.
      \item Look for all global root labels that are not present in the current
            hyperplane and have no right neighbour; delete them and send the
            number of sites to the left neighbour.
      \item When a global label is informed that its right neighbour was
            deleted, and it has no left neighbour, convert it to local.
      \end{enumerate}
\item \emph{Counting}:
      \begin{enumerate}
      \item Count local clusters.
      \item Concentrate global clusters.
      \item Count global clusters.
      \item Look for a global cluster which has not been concentrated. If
            it exists, we have connectivity. Sum up this cluster explicitly.
      \item Do output.
      \end{enumerate}
\end{enumerate}

\section{Results of Monte Carlo Studies}
\noindent
All simulations were done right at the critical threshold $p_c$ (except those
for Fig.~2) and on
square ($2d$), simple cubic ($3d$), and simple hypercubic ($4d$) lattices.
The values for $p_c$ were taken from literature:

\begin{tabular}{lll}
$2d$ & $p_c=0.5927464$ & cf.~\cite{pc2d}\\
$3d$ & $p_c=0.311608$  & cf.~\cite{pc3d}\\
$4d$ & $p_c=0.196889$  & cf.~\cite{pc4d}
\end{tabular}

As boundary conditions, one direction was busbar in two dimensions resp.
open in higher dimensions, one direction was periodic, and for more than
two dimensions all other directions were helical. This mix came quite natural
during implementing the parallel algorithm.

\subsection{Cluster size distribution}
\noindent
For the number $n_s$ of clusters of size $s$ in our system, we expect
a distribution of kind $n_s \propto s^{-\tau}$
right at the critical threshold $p_c$.
To make analysis easier, we look at $N_s$, the number of clusters with at least
$s$ sites:
\begin{equation}
N_s = \sum_{s'=s}^{\infty} n_{s'} 
\simeq k_0 s^{-\tau+1}\,.
\end{equation}

In a double-logarithmic plot, we would expect a straight line with slope
$-\tau+1$. Deviations from the power law would not be visible due to the
logarithmic scale, thus we plot $s^{\tau-1}N_s$ linearly on the $y$-axis.
We can clearly see in Fig.~\ref{fig_cluster} the deviation for small $s$ (due to corrections
to scaling, see below) and for large $s$ (due to boundary conditions).

\begin{figure}[htbp]
\label{fig_cluster}
\begin{center}
\rotatebox{270}{\includegraphics[height=100mm]{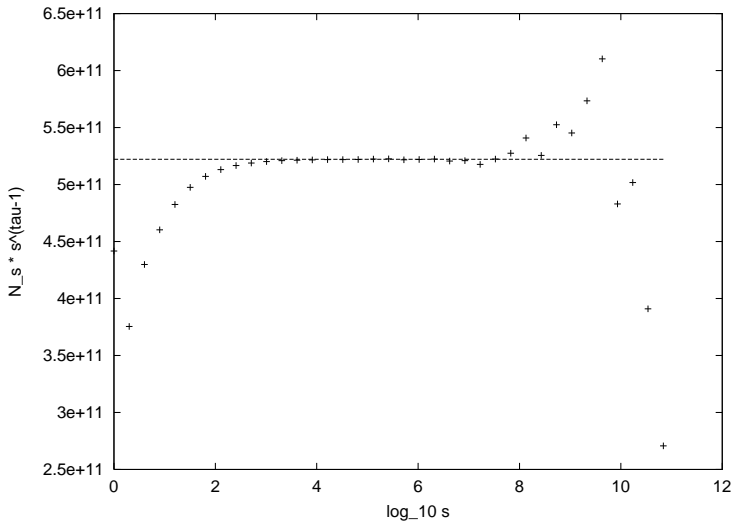}}
\caption{Cluster size distribution for $L=4000256$ in two dimensions. The
dashed line corresponds to the theoretically asymptotic behaviour.}
\end{center}
\end{figure}

The values found for the proportionality constant $k_0$ (normalized by number
of lattice sites) and the exponent $\tau$ are:

\begin{tabular}{lll}
$2d$ & $k_0=0.032627(6)$ & $\tau=187/91$\footnotemark \\
$3d$ & $k_0=0.057423(3)$ & $\tau=2.190(2)$\\
$4d$ & $k_0=0.0611(4)$ & $\tau=2.313(2)$
\end{tabular}
\footnotetext{the value for $\tau$ is supposed to be known exactly}

\begin{figure}[htbp]
\label{fig_largest}
\begin{center}
\rotatebox{270}{\includegraphics[height=100mm]{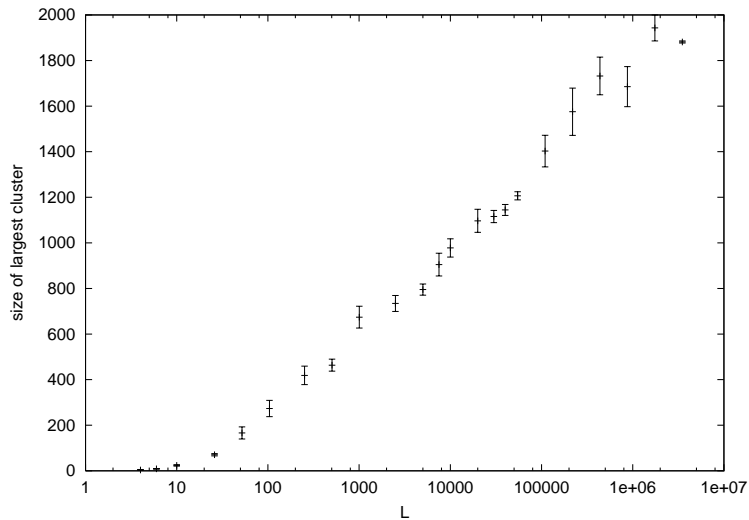}}
\caption{Average size of the largest cluster versus system size $L$ at
$p=0.5$ in two dimensions. The error bars represent the statistical error.
Number of runs done
for each $L$: eight for $L < 200 \, \mbox{k}$, four for
$200 \, \mbox{k} < L < 1 \, \mbox{M}$, two for $L>1 \, \mbox{M}$.}
\end{center}
\end{figure}

Below $p_c$, we expect the size of the largest cluster to be proportional
to $\log (L)$, cf.~Fig.~\ref{fig_largest}.

\subsection{Corrections to scaling}
\noindent
For small $s$, the clusters do not follow the power-law
$n_s \propto s^{-\tau}$;
the system is not self-similar under renormalization, as the lattice spacing
is an inherent length, very visible for small clusters. Thus our power-law
is modified: $n_s = k_0 s^{-\tau} (1-k_1 s^{-\Delta_1})$. We find the
exponent $\Delta_1$ (and the factor $k_1$) by plotting 
$\log ( N_s / (k_0 s^{-\tau+1}) - 1 )$ against $\log s$; for carefully
adjusted $k_0$ and $\tau$, this yields a straight line with slope $\Delta_1$
and intercept $k_1$.

\begin{figure}[htbp]
\begin{center}
\label{fig_corr}
\rotatebox{270}{\includegraphics[height=100mm]{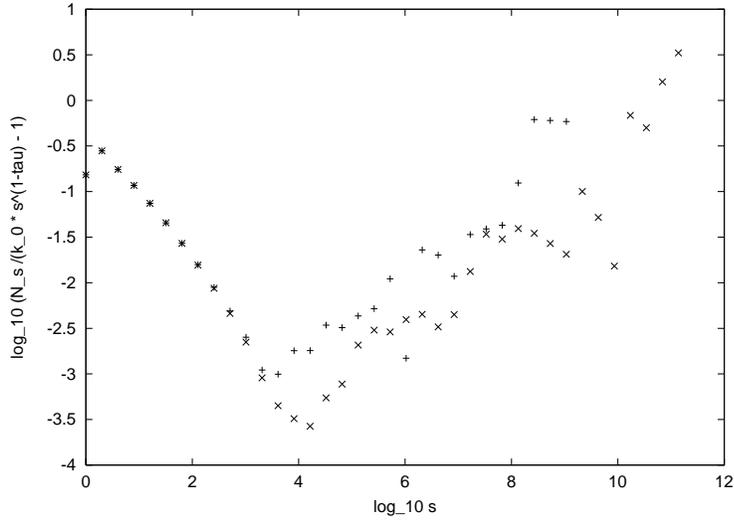}}
\caption{Corrections to scaling in two dimensions, for $L=500 {\rm k}$ ($+$)
and $L=4 {\rm M}$ ($\times$). The larger system offers higher precision, as the
finite-size effects are weaker.}
\end{center}
\end{figure}

We get a straight line only for
small and intermediate $s$. For large $s$, the finite-size effects increase
the $n_s$ over the expected value and limit the precision, cf.~Fig.~\ref{fig_corr}.
Implementing
fully periodic boundary conditions could give an improvement in precision.

The values found for the corrections to scaling amplitude and exponent are: 

\begin{tabular}{lll}
$2d$ & $k_1=0.50(2)$ & $\Delta_1=0.70(2)$\\
$3d$ & $k_1=0.48(5)$ & $\Delta_1=0.60(8)$\\
$4d$ & $k_1=0.5(1)$  & $\Delta_1=0.5(1)$\\
\end{tabular}

\subsection{Number density}
\noindent
The total number of clusters divided by the number of sites in the lattice is
called the number density, right at $p_c$ the critical number density. For
small systems, we expect finite-size corrections proportional to $1/L$.
Statistical fluctuations depend on the total number of sites in the lattice
and are proportional to $1/L^d$. Thus finite-size effects can be seen well
in higher dimensions.

\begin{figure}[htbp]
\label{fig_nc}
\begin{center}
\rotatebox{270}{\includegraphics[height=100mm]{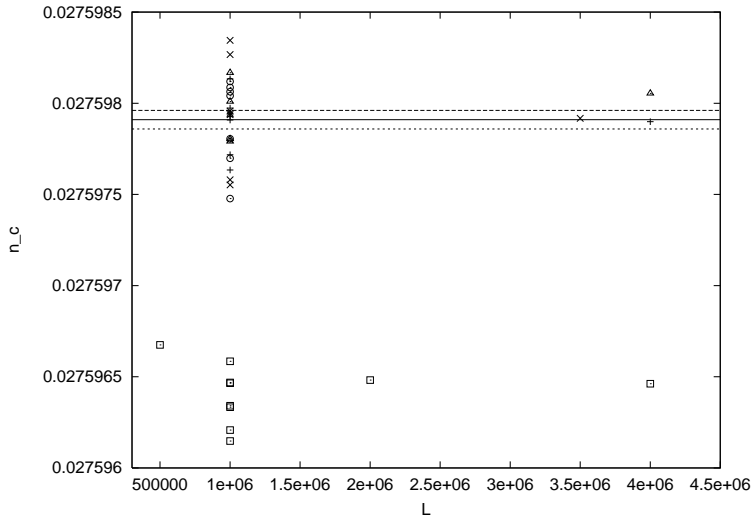}}
\caption{Number densities at $p_c$ for various system sizes $L$ in two
dimensions, using different PRNGs: R(103,250) ($+$), R(471,1586,6988,9689)
($\times$),
ibm*16807 (boxes), R(18,36,37,71,89,124) (circles), and ibm*$13^{13}$
(triangles). The solid line corresponds to the average for the $L=1 {\rm M}$
runs of the two-tap, four-tap, and six-tap R(\dots) generators, the dashed
lines to the statistical error.}
\end{center}
\end{figure}

For two dimensions, we can simulate large $L$ and have no visible finite-size
effects.
But we can clearly see a dependence on the used random number generator. The
classic and simple {\tt ibm = ibm * 16807} produces wrong results, whereas
others agree well with each other (i.~e.~R(103,250), R(471,1586,6988,9689),
R(18,36,37, 71,89,124),$^{11}$ even {\tt ibm = ibm * 13**13} seems to be
acceptable). Thus the
number density can serve as a test for bad random numbers, cf.~Fig.~\ref{fig_nc}.

The values found for the number density are:

\begin{tabular}{ll}
$2d$ & $n_c=0.02759791(5)$\\
$3d$ & $n_c=0.0524387(3)$\\
$4d$ & $n_c=0.0519980(2)$\\
\end{tabular}

\section{Efficiency}
\noindent
When implementing an algorithm on a parallel computer, we want to know how
efficient our implementation is, i.~e.~if not too much time is wasted for
communication.

The amount of neccessary communication is roughly proportional to the interface
between the strips, i.~e.~$NL^{d-1}$. This communication is the main difference
between the sequential and parallel implementation, and reduces the parallel
efficiency.

In two dimensions, a sequential implementation is roughly as fast as the
parallel one, which means that the effort for communication is neglectable.
In four dimensions, the fraction of runtime needed for communication is
significant, but the time for the local part is still larger.

Although the domain decomposition in perpendicular strips requires more
communication than other methods, it is still efficient enough for up to
four dimensions. For even higher dimensions, this may change.

\section{Summary}
\noindent
Parallelizing the Hoshen-Kopelman by dividing the lattice into strips
perpendicular to the hyperplane of investigation is a viable approach.
Using this method, world record simulations in two, three, and four dimensions
have been carried out, beating the old world records.$^{4,5}$ High precision results
for the Fisher exponent
$\tau$, the corrections to scaling exponent $\Delta_1$, and the number
density $n_c$ have been obtained. The values found are in reasonable agreement
with those from other groups.$^{4,5,9}$

\section*{Acknowledgements}
\noindent
Fruitful discussions with D.~Stauffer, N.~Jan, and R.~Ziff are gratefully
acknowledged. I would also like to thank the Research Center J\"ulich for
computing time on their Cray T3E.

\end{document}